\documentclass[journal]{IEEEtran}
\usepackage{amsmath,amsfonts}
\usepackage{algorithmic}
\usepackage{algorithm}
\usepackage{array}
\usepackage[caption=false,font=normalsize,labelfont=sf,textfont=sf]{subfig}
\usepackage{textcomp}
\usepackage{stfloats}
\usepackage{url}
\usepackage{verbatim}
\usepackage{graphicx}
\usepackage{cite}
\usepackage{booktabs}
\usepackage{siunitx}
\usepackage{tikz}
\usepackage{listings}
\usetikzlibrary{arrows.meta,positioning}
\usepackage{hyperref}
\usepackage{orcidlink}

\DeclareSIUnit\months{months}

\tikzset{
  >=Latex,
  edge/.style={->,thick},
  lbl/.style={midway,fill=black!3,inner sep=1pt,font=\tiny},
  circ/.style={circle, draw=black, fill=none, minimum size=24mm, align=center},
  identity/.style={circ},
  email/.style={circ},
  ghuser/.style={circ},
  ghrepo/.style={circ},
  grepo/.style={circ},
  gcommit/.style={circ},
  ghcommit/.style={circ}
}

\begin{document}

\title{Beneath the Mask: Can Contribution Data Unveil Malicious Personas in Open-Source Projects?}

\author{
    Ruby Nealon\orcidlink{0009-0003-7432-6943}
    \thanks{This work has been submitted to the IEEE for possible publication. Copyright may be transferred without notice, after which this version may no longer be accessible.}
}

\maketitle

\begin{abstract}
In February 2024, after building trust over two years with project maintainers by making a significant volume of legitimate contributions, GitHub user ``JiaT75''  self-merged a version of the XZ Utils project containing a highly sophisticated, well-disguised backdoor targeting \textit{sshd} processes running on systems with the backdoored package installed. A month later, this package began to be distributed with popular Linux distributions until a Microsoft employee discovered the backdoor while investigating how a recent system upgrade impacted the performance of SSH authentication. Despite its potential global impact, no tooling exists for monitoring and identifying anomalous behavior by personas contributing to other open-source projects. This paper demonstrates how Open Source Intelligence (OSINT) data gathered from GitHub contributions, analyzed using graph databases and graph theory, can efficiently identify anomalous behaviors exhibited by the ``JiaT75'' persona across other open-source projects.
\end{abstract}


\section{Introduction}
\IEEEPARstart{O}{n} February 23, 2024, Version 5.6.0 of the XZ Utils project \textit{libxz}\slash\textit{liblzma} library package was released, introducing a backdoor targeting the \textit{sshd} remote login server. Just over a month later, on March 29, it was discovered by Andres Freund, a software engineer at Microsoft, who first disclosed it to the Openwall mailing list. The backdoor was introduced to the project source code by a contributor known as Jia Tan (or ``JiaT75''). JiaT75 had first started contributing to the project in January 2022, engaging in an incredibly sophisticated campaign of sock-puppetry. By the time the backdoored version was released, they were recognized as a project member with commit rights, publishing the backdoored version without prior review by other maintainers\cite{James2024}.

The backdoor received much attention after the disclosure for reasons relating to the discovery of the backdoor, the backdoor itself, and the extensive time and effort the attacker invested in building trust with the ``JiaT75'' persona. In his own words, Freund, who disclosed the backdoor, did not consider himself a security researcher; instead, he discovered the backdoor whilst investigating high CPU usage on SSH logins, which he initially believed to be a performance regression \cite{Freund2024}.

 The choice of the \textit{sshd} package as a target was also interesting, as \textit{sshd} does not depend directly or indirectly on the backdoored package. However, many major Linux distributions, like Ubuntu and Debian, use \textit{systemd} for service management. They distribute versions of packages like \textit{sshd} patched for compatibility. Because \textit{systemd} depends on XZ Utils \textit{liblzma} package, the backdoored library was loaded indirectly with the \textit{sshd} process \cite{Freund2024}. 

For many, though, the most interesting part of the backdoor was the ``JiaT75'' persona and the level of trust over time the attacker had built by making significant legitimate-seeming contributions to the project \textendash{} most of which are still present in the source code. Existing timelines and analyses \cite{Tumbleson2024, Boehs2024} demonstrate that the social engineering done by the attacker behind the backdoor likely extends far beyond the XZ Utils GitHub repository and even the specific ``Jiat75'' persona itself. 

In the aftermath of the backdoor, work has already been done to implement technical controls against the weaknesses in how open-source packages are released and distributed that were exploited by the attacker. For example, ``backseat-signed'' is a tool created in response to the backdoor to assist with verifying the cryptographic chain-of-custody of Linux distribution packages and their upstream source code inputs \cite{Kpcyrd}. However, despite a lot of the social engineering identified in previously mentioned writing taking place within the structures of the GitHub and Git repositories themselves, there have been no attempts so far to automate detecting the anomalous behaviors exhibited by the JiaT75 persona in other open-source project repositories. 

\subsection{Graph Databases}

Databases used by applications requiring data retrieval and storage are typically ``relational''; data is modeled and stored in structured tables with rows and columns. This is the case for small databases aimed at personal and small business use, such as Microsoft Access, and commercial and non-commercial large-scale offerings like MySQL, Microsoft SQL Server, and Oracle. In contrast, graph databases model and store data as nodes connected by relationships (also referred to as edges or links). Though graph data can be modeled and stored in relational databases, graph databases are generally more performant for queries that traverse graph data. Graph database query languages are also considered more straightforward when writing queries against graph data \cite{VicknairEtAl}.

Cybersecurity tools already use graph databases for use cases involving analyzing and investigating relationship-dense data. A well-known example of this is BloodHound. Using Active Directory objects like users and groups and their relationships like memberships and privileges, BloodHound uses graph theory to find possible paths for privilege escalation from a given level of access in an environment, as well as the discovery of unusual or unexpected relationships \cite{SpecterOps}.

The natural provenance inside Git repositories, spread across many projects and contributors, makes data from source forge websites like GitHub inherently relationship-dense. This creates an opportunity for using graph theory and graph databases to investigate and query open-source project contribution data. 

\section{Research Method}

To validate the utility of signals from OSINT contributor data, public data was gathered from XZ Utils and a control group of similar projects. The data was collected from both GitHub and the GitHub-hosted Git repository itself, and ingested into a graph database. Behaviors anomalous to typical open-source contributors identified in existing writing about the backdoor were analyzed, quantified, and written as graph database queries. The queries were then executed against the set of data collected from the scoped projects, and their results discussed for the suitability of the query in discovering behavior that may be worthy of further scrutiny.

\subsection{Scoping}

A total of 19 repositories hosted on GitHub.com containing the source code for direct and indirect, required, optional, and \textit{make} dependency packages of \textit{systemd} were selected from the Arch Linux package registry. The selected projects range from small (such as \textit{hwdata} with 414 commits from 14 contributors) to large (\textit{libarchive} with 6640 commits from 307 contributors) with varying levels of organization and governance. The projects selected also have varying levels of ``popularity'' (presence across sampled, opted-in Arch Linux installations) per the archlinux.de ``Pkgstats'' project \cite{Archlinuxde}.

\begin{table*}
	\caption{Projects in Scope for Collection (Data as of March 27, 2025)}
	\centering
	\begin{tabular}{
        l
        l
        S[table-format=2.2]
        S[table-format=4.0]
        S[table-format=4.0]
    }
		\toprule
		{GitHub repository} & {Arch package name} & {Arch package popularity (\%)} & {\# of commits in main branch} & {\# of unique contributors}
        \\\hline 
		\midrule
		{tukaani-project/xz} & {core/xz} & 100\%  & 2854 & 29\\\hline 
		{linux-audit/audit-userspace} & {core/audit} & 100\%  & 2725 & 62\\\hline 
		{p11-glue/p11-kit} & {core/libp11-kit} & 100\%  & 1474 & 82\\\hline 
		{seccomp/libseccomp} & {core/libseccomp} & 100\%  & 1047 & 66\\\hline 
		{vcrhonek/hwdata} & {core/hwdata} & 99.73\%  & 414 & 14\\\hline 
		{besser82/libxcrypt} & {core/libxcrypt} & 99.97\%  & 830 & 23\\\hline 
		{lz4/lz4} & {core/lz4} & 100\%  & 3615 & 177\\\hline 
		{fukuchi/libqrencode} & {extra/qrencode} & 69.97\%  & 807 & 29\\\hline 
		{thom311/libnl} & {core/libnl} & 100\%  & 2138 & 143\\\hline 
		{tpm2-software/tpm2-tss} & {core/tpm2-tss} & 99.55\%  & 2905 & 119\\\hline 
		{libpwquality/libpwquality} & {extra/libpwquality} & 41.11\%  & 263 & 57\\\hline 
		{PJK/libcbor} & {extra/libcbor} & 44.29\%  & 1379 & 40\\\hline 
		{libexpat/libexpat} & {core/expat} & 100\%  & 4501 & 89\\\hline 
		{json-c/json-c} & {core/json-c} & 100\%  & 1376 & 139\\\hline 
		{libarchive/libarchive} & {core/libarchive} & 100\%  & 6640 & 307\\\hline 
		{PCRE2Project/pcre2} & {core/pcre2} & 100\%  & 2064 & 58\\\hline 
		{eliben/pyelftools} & {extra/python-pyelftools} & 23.66\%  & 716 & 94\\\hline 
		{apjanke/ronn-ng} & {extra/ruby-ronn-ng} & 1.7\%  & 591 & 22\\\hline 
		{xkbcommon/libxkbcommon} & {extra/libxkbcommon} & 89.8\%  & 2510 & 66\\\hline 
		{shadow-maint/shadow} & {core/shadow} & 99.42\%  & 3865 & 177\\\hline 
		\bottomrule
	\end{tabular}
	\label{tab:scopedprojects}
\end{table*}

\par As data was to be collected for the GitHub repositories and their related entities, the GitHub GraphQL API was used. GraphQL is a schema and query language that allows for requesting objects and their fields as graphs, allowing for more efficient API queries returning only the required data \cite{Vadlamani2021}. The GitHub GraphQL API enforces an hourly point-based rate limit system, scoring requests based on the total number of entities their query loads \cite{GitHubRateLimit}. As it can take several hours to load all relevant data for some larger repositories, the number of projects scoped was intentionally limited for this experiment.

\subsection{Tool development and environment}

\subsubsection{Environment}

The Neo4j graph database  was used for this experiment, the same used by the previously mentioned BloodHound tool. The database and tooling used were run on a single macOS machine, with all data collected with GitHub GraphQL API and \texttt{git clone} calls.

\subsubsection{Data Collection}

To collect data efficiently and within the constraints of the aforementioned limits of the GitHub GraphQL API, automation was developed as a ``gem'' using the Ruby programming language. The automation first queried the total number of nodes for the desired connections of each repository (for example, pull requests or issues), and used GraphQL variables to control the pagination cursor and batch size. This minimized the number of requests and API quota used and ensured that only new, relevant data was loaded. Once all data was loaded, the raw paginated connections were joined into arrays and serialized into JSON data files for flexibility with schema changes during data ingestion.

\subsubsection{Data Ingestion}

The same gem defines ORM models for nodes representing the different entities using ActiveGraph \textendash{} a Ruby gem providing similar functionality to the popular ActiveRecord ORM distributed as a part of the Ruby on Rails web framework \cite{Neo4JRb}. The collected data was traversed as a tree, upserted into nodes, and mapped into a temporary list of all possible relationships between nodes. Once all nodes were inserted, the temporary list of relationships was then reduced to the minimal number required, removed of duplicates, and inserted in bulk.

After the data queried from the GitHub GraphQL API was ingested, the branch, commit, and author/committer identity data were also ingested from the Git repository retrieved with the bare \texttt{git clone} call. 

Finally, the minimal number of additional GitHub GraphQL API calls was made to resolve additional GitHub user data associated with email addresses present in the loaded Git commits.

A schema of node labels and relationships was defined via the ActiveGraph ORM to mirror the same objects and hierarchies found in each external data source. For example, the \textit{HAS\_GITHUB\_PULL\_REQUEST} relationship from \textit{GithubRepository} to \textit{GithubPullRequest} mirrors how pull requests are loaded for a repository in the API call. 

The final schema for all node labels and their defined external data sources is described in Table~\ref{tab:nodelabels} below. 

\begin{table}
	\caption{Node Labels and Their External Data Sources}
	\centering
	\begin{tabular}{ll}
		\toprule
		Node label & Data source\\\hline
		\midrule
		GithubCommit & GitHub GraphQL API\\\hline 
		GithubCommitComment & GitHub GraphQL API\\\hline 
		GithubDiscussion & GitHub GraphQL API\\\hline 
		GithubDiscussionComment & GitHub GraphQL API\\\hline 
		GithubIssue & GitHub GraphQL API\\\hline 
		GithubIssueComment & GitHub GraphQL API\\\hline 
		GithubOrganization & GitHub GraphQL API\\\hline 
		GithubPullRequest & GitHub GraphQL API\\\hline 
		GithubPullRequestReview & GitHub GraphQL API\\\hline 
		GithubPullRequestReviewComment & GitHub GraphQL API\\\hline 
		GithubRepository & GitHub GraphQL API\\\hline 
		GithubUser & GitHub GraphQL API\\\hline 
		GithubUserContentEdit & GitHub GraphQL API\\\hline 
		GitBranch & Git repository data\\\hline 
		GitCommit & Git repository data\\\hline 
		GitIdentity & Git repository data\\\hline 
		\bottomrule
	\end{tabular}
	\label{tab:nodelabels}
\end{table}

\subsection{Defining Anomalous Criteria and Method of Investigation}

\subsubsection{Review of Existing Analyses and Writing}

Existing analyses and articles covering the extended timeline leading up to the insertion and discovery of the XZ Utils backdoor cover social engineering efforts far beyond what would be detectable from just the GitHub and Git repository data. However, consistent patterns are identified regarding the activity of the JiaT75 user in the project.

The ''JiaT75'' GitHub user was created in January 2021, more than 3 years before the backdoor was created. However, the Git repository for XZ Utils dates back as far as 2007, and the project dates back even further. After creating their GitHub user, they immediately began contributing to other projects, including \textit{libarchive} \textendash{} one of the scoped repositories for data collection. In April 2022, the Jia Tan/JiaT75 persona begins regularly contributing to the XZ Utils project, and in January 2023, they self-merge their first pull request.

While contributing to multiple projects and new maintainers joining projects is not unusual, the frequency and pacing of the JiaT75 persona are. Connor Tumbleson, a software engineer, highlighted this in his blog post covering the attack situation:

\begin{quote}
    \par So now I was curious when Jia Tan was created on GitHub and scrolled all the way back.
    {\ldots}
    \par This user was not a very old account being only created in January of 2021 and one of their first public merged pull requests is also under investigation.
    \par I scroll the user's history on GitHub and I'm blown away - commits and fixes to projects all over - some seemingly great and normal while others being reverted as quick as possible. \cite{Tumbleson2024}
\end{quote}

The short time to become an active contributor and gain trust within the project is also noted. Evan Boehs, a software engineer also blogging about the attack, writes:

\begin{quote}
    \par Three days after the emails pressuring Lasse Collin to add another maintainer, JiaT75 makes their first commit to xz: Tests: Created tests for hardware functions... Since this commit, they become a regular contributor to xz (they are currently the second most active). 
    {\ldots}
    \par JiaT75 merges their first commit on January 7, 2023, which gives us a good indication of when they fully gain trust. \cite{Boehs2024}
\end{quote}

\subsubsection{Quantifiable Anomalous Criteria}

From these observations, several quantifiable factors are identified that, in combination, could suggest anomalous behavior:

\begin{itemize}
\item Project history age (the earliest and latest authored date of any commit in a project)
\item Contribution activity within a project (the percentage of commits authored by a contributor in a project)
\item Contribution history within a project (the earliest and latest dates of commits authored by the contributor)
\item History of self-merging pull requests without review within a project (if any, the earliest merge request by a contributor merged without review by another party)
\end{itemize}

\subsubsection{Defining the Sets of Anomalous Criteria To Investigate}

This research investigated two sets of criteria that indicate anomalous behavior: (1) contributors self-merging pull requests without review with limited project involvement and (2) contributors making an unusually significant share of all-time contributions for their total presence in the repository history.

Each set of criteria was investigated by first writing and executing an unscoped query, analyzing the results and defining thresholds/filters where appropriate, then re-executing and discussing the set of criteria and results with regard to their suitability as a signal of potential social engineering. With the exception of the JiaT75 persona, contributor usernames are anonymized in the results.

\section{Findings and Discussion}

\subsection{Enriching Collected Data With Additional Relationships}

While the data collected from the GitHub GraphQL API and bare Git repository already have inherent hierarchies that accurately describe the logical relationships between entities, like the repository having pull requests example mentioned earlier, other relationships valuable for investigation were not present in the source data. 

To assist with expressively writing queries to look for deep relationships in the data set, the following additional relationships were defined and populated for the ingested data.

\subsubsection{IS\_GIT\_REPOSITORY and IS\_GIT\_COMMIT}

These relationships link a \textit{GithubRepository} or \textit{GithubCommit} loaded from the GraphQL API to its corresponding \textit{GitRepository} or \textit{GitCommit} loaded from the bare Git repository data. 

\begin{figure*}[t]
    \centering
    \begin{lstlisting}[frame=single,framexleftmargin=2mm,framextopmargin=3mm,framexbottommargin=3mm]
    MATCH (ghr:GithubRepository)-[:IS_GIT_REPOSITORY]->(gr:GitRepository),
          (ghc:GithubCommit)-[:IS_GIT_COMMIT]->(gc:GitCommit)<-[:HAS_GIT_COMMIT]-(gr:GitRepository)
    RETURN * LIMIT 1;
    \end{lstlisting}
    \begin{tikzpicture}[node distance=3cm,every node/.style={font=\footnotesize}]
      \node[ghrepo]   (ghr) at (-7.5,0) {tukaani-project/xz};
      \node[grepo]    (gr)  at (-2.5,0) {https://github.com/\ldots};
      \node[gcommit]  (gc)  at (2.5,0) {ef652a\ldots};
      \node[ghcommit] (ghc) at (7.5,0) {ef652a\ldots};
    
      \draw[edge] (ghr) -- node[lbl,above]{IS\_GIT\_REPOSITORY} (gr);
      \draw[edge] (gr)  -- node[lbl,above]{HAS\_GIT\_COMMIT}   (gc);
      \draw[edge] (ghc) -- node[lbl,above]{IS\_GIT\_COMMIT}    (gc);
    \end{tikzpicture}
	\caption{Query Demonstrating \textit{IS\_GIT\_REPOSITORY} and \textit{IS\_GIT\_COMMIT}}
	\label{fig:gitrepocommitquery}
\end{figure*}
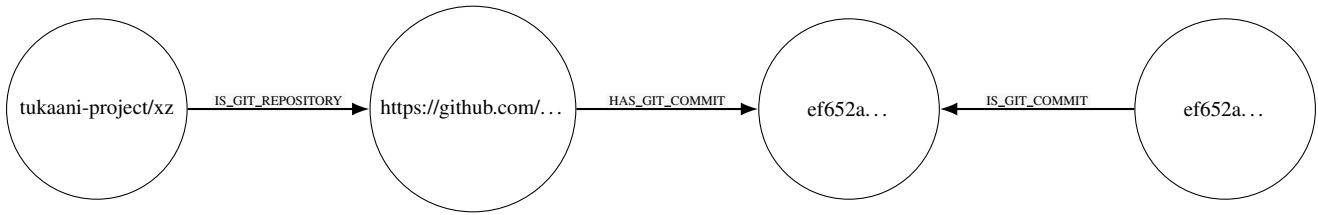

An example query demonstrating these relationships run against the scoped data set is shown in Figure~\ref{fig:gitrepocommitquery}.

\subsubsection{LINKED\_TO\_GITHUB\_USER}

This relationship links a \textit{GitIdentity} to a \textit{GithubUser} from the set of commits collected for a repository from the GraphQL API. As GitHub links a commit to a user by the email address, only a single commit needs to be loaded from the GraphQL API, where the email address is the committer or author to create this relationship between the resulting \textit{GithubUser} node and all \textit{GitIdentity} nodes with the same email address.

\begin{figure*}[t]
    \centering
    \begin{lstlisting}[frame=single,framexleftmargin=2mm,framextopmargin=3mm,framexbottommargin=3mm]
    MATCH (gi:GitIdentity)-[:LINKED_TO_GITHUB_USER]->(ghu:GithubUser)
    WHERE ghu.login = "JiaT75"
    RETURN *;
    \end{lstlisting}
    \begin{tikzpicture}[node distance=3cm,every node/.style={font=\footnotesize}]
      \node[ghuser]   (gh) at (0,0) {JiaT75};
      \node[identity] (a)  at (0,4) {jiat75};
      \node[identity] (b)  at (-6,0) {Jia Tan};
      \node[identity] (c)  at (6,0) {Jia Cheong Tan};
    
      \draw[edge] (a) -- node[lbl,right]{LINKED\_TO\_GITHUB\_USER} (gh);
      \draw[edge] (b) -- node[lbl,above]{LINKED\_TO\_GITHUB\_USER} (gh);
      \draw[edge] (c) -- node[lbl,above]{LINKED\_TO\_GITHUB\_USER} (gh);
    \end{tikzpicture}
	\caption{Query Demonstrating \textit{LINKED\_TO\_GITHUB\_USER}}
	\label{fig:linkedgithubquery}
\end{figure*}
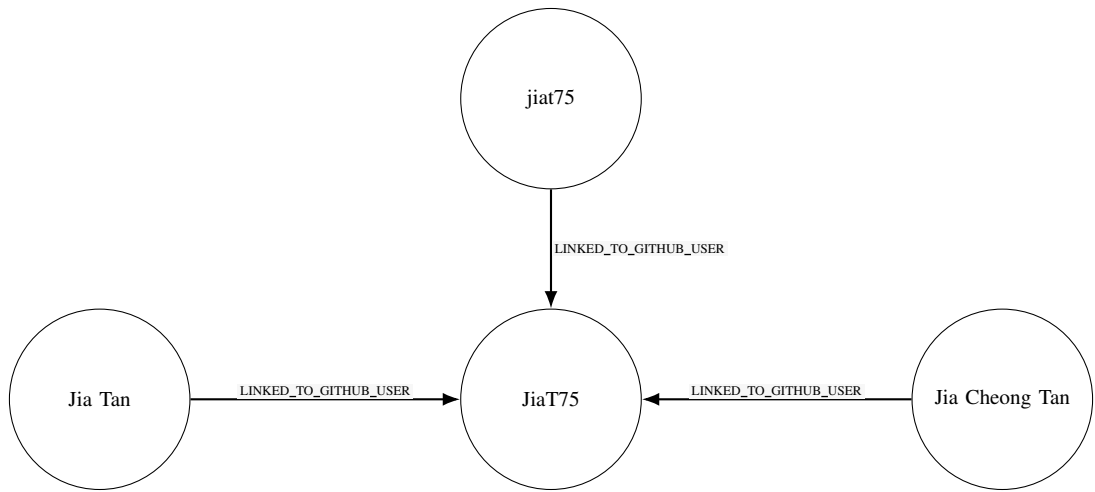

An example query demonstrating this relationship by loading the \textit{GitIdentity} nodes linked to the \textit{GithubUser} node for ``JiaT75'' is shown in Figure~\ref{fig:linkedgithubquery}.

\subsubsection{HAS\_EMAIL}

This relationship links a GithubUser, GithubOrganization, or GitIdentity to a new \textit{Email} node created for each unique email address discovered.

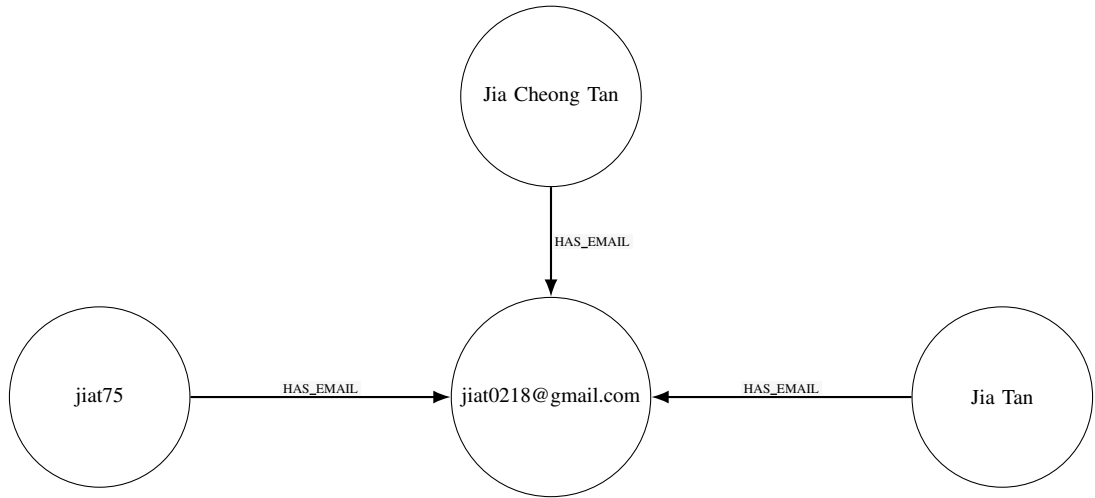
\begin{figure*}[t]
    \centering
    \begin{lstlisting}[frame=single,framexleftmargin=2mm,framextopmargin=3mm,framexbottommargin=3mm]
    MATCH (gi:GitIdentity)-[:HAS_EMAIL]->(e:Email)
    WHERE e.address = "jiat0218@gmail.com"
    RETURN *;
    \end{lstlisting}
    \begin{tikzpicture}[node distance=3cm,every node/.style={font=\footnotesize}]
      \node[email]    (e)   at (0,0) {jiat0218@gmail.com};
      \node[identity] (jct) at (0,4) {Jia Cheong Tan};
      \node[identity] (u1)  at (-6,0) {jiat75};
      \node[identity] (jt)  at (6,0) {Jia Tan};
    
      \draw[edge] (jct) -- node[lbl,right]{HAS\_EMAIL} (e);
      \draw[edge] (u1)  -- node[lbl,above]{HAS\_EMAIL} (e);
      \draw[edge] (jt)  -- node[lbl,above]{HAS\_EMAIL} (e);
    \end{tikzpicture}
	\caption{Query Demonstrating \textit{HAS\_EMAIL}}
	\label{fig:hasemailquery}
\end{figure*}

An example query demonstrating this relationship by loading the \textit{GitIdentity} nodes linked to the \textit{Email} node for the email address ``jiat0218@gmail.com'' is shown in Figure ~\ref{fig:hasemailquery}.

\subsection{Querying for Unreviewed, Self-Merged Pull Requests}

\subsubsection{Query Creation}

The first candidate criteria set to investigate earlier was a contributor’s history of unreviewed (by another party), self-merged pull requests in relation to the age of the repository and when they first began contributing to it. To effectively measure this, a query was prepared collecting the following values for each \textit{GithubUser} in each \textit{GithubRepository} where they had self-merged at least one pull request:

\begin{itemize}
\item The \textit{GitHubUser} username, creation timestamp, first authored commit timestamp, and all associated email addresses
\item The \textit{GithubRepository} name with its owner, creation timestamp, and first-authored commit timestamp
\item The URL of and merge timestamp of the first self-merged pull request
\item The total number of self-merged pull requests
\item The difference between the \textit{GithubUser} and \textit{GithubRepository} first authored commit timestamps (i.e., the repository age at the time of the first contribution)
\item The difference between the \textit{GithubUser}'s first authored commit timestamp and the merge timestamp of the first self-merged pull request
\end{itemize}

This query returned 40 users, approximately 2\% of the 1996 unique users in the data set who had authored a pull request. No user in the data set had self-merged an unreviewed pull request in more than one repository.

The data returned for the JiaT75 persona differs slightly from what Boehs \cite{Boehs2024} stated in his timeline but was confirmed as correct by cross-referencing against GitHub and the Git repository; the first authored commit by JiaT75 in the repository was in late January 2022 (though not \textit{committed} until July), and their first self-merged pull request was in December of the same year. GitHub does not show separate dates for authored and committed times of commits, so this could have caused a misinterpretation.

\begin{figure}
	\centering
	\includegraphics[width=0.48\textwidth]{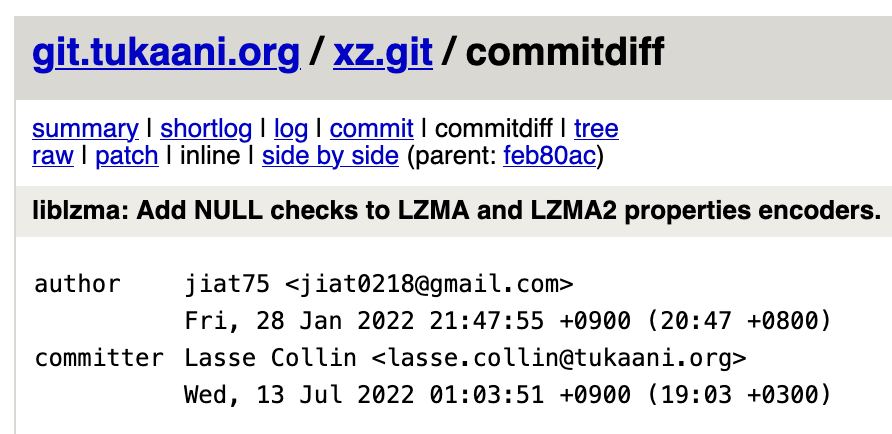}
	\caption{JiaT75’s First Authored Commit to XZ Utils}
	\label{fig:jiatanfirstcommit}
\end{figure}

\begin{figure}
	\centering
	\includegraphics[width=0.48\textwidth]{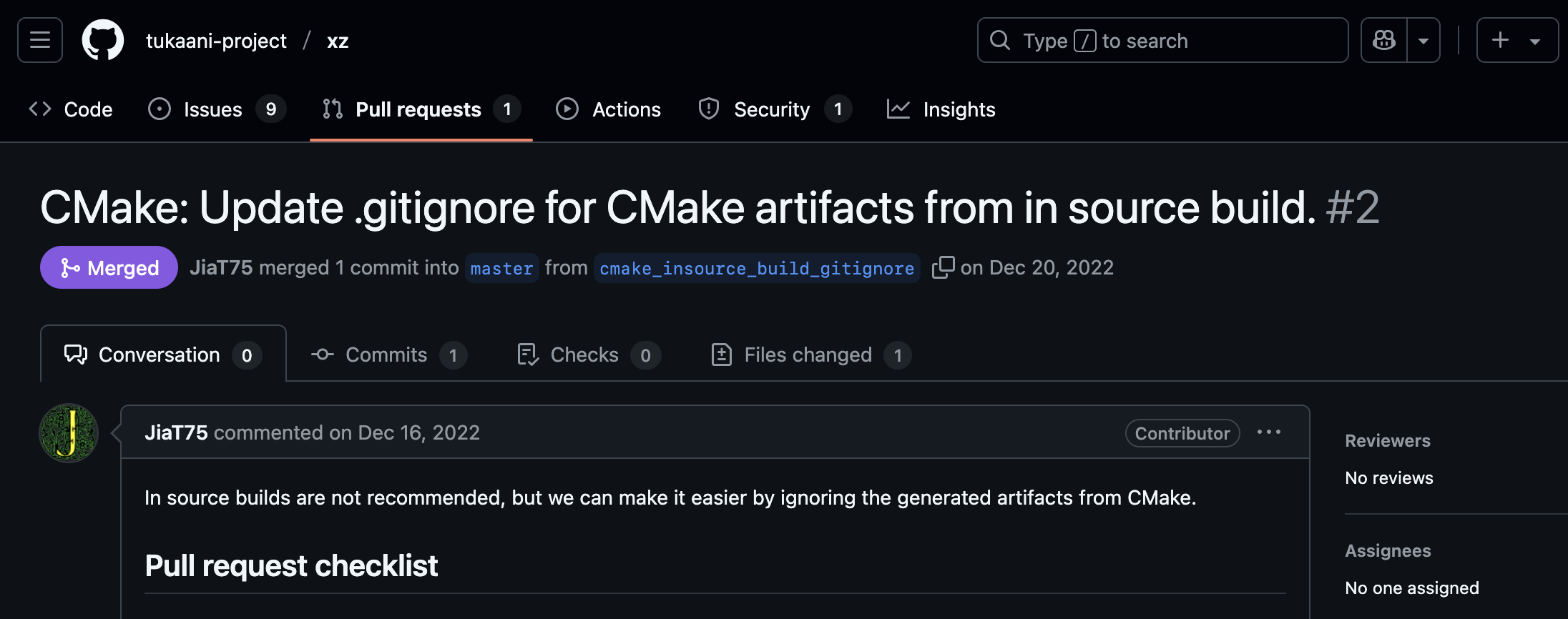}
	\caption{JiaT75’s First Unreviewed, Self-Merged Pull Request}
	\label{fig:jiatanfirstselfmerge}
\end{figure}

The correct contributor age for JiaT75, confirmed from the dates in Figures~\ref{fig:jiatanfirstcommit} and~\ref{fig:jiatanfirstselfmerge}, returned in the results, is 10 months. Nearly half of the results (19) have a contributor age of less than 24 months, which was used as a threshold for the final query to exclude long-term project participants.

The query was then further iterated, adding exclusions for irrelevant (i.e., now long-term contributors) and edge cases:

\begin{itemize}
\item Excluded results where the user’s first authored Git commit was more than 4 years ago \textendash{} reduced the results from 19 to 7 records
\item Excluded results where the contributor age was less than or equal to zero \textendash{} reduced the results from 7 records to 4 records
\end{itemize}

\subsubsection{Discussion of Results}

Table~\ref{tab:resultsunreviewedselfmerged} presents the relevant data from the results after executing the final query. The full query used is available in the supplementary material.

\begin{table*}
	\caption{Results of the Unreviewed, Self-Merged Pull Request Query}
	\centering
	\begin{tabular}{
        l
        l
        S[table-format=2.0]
        S[table-format=2.0]
    }
		\toprule
		Name & Repository & {Contributor age at first self-merge} & {\# of unreviewed, self-merged pull requests}
        \\\hline
		\midrule
        Contributor A & PCRE2Project/pcre2 & \SI{2}{\months} & 43\\\hline
        Contributor B & PCRE2Project/pcre2 & \SI{7}{\months} & 7\\\hline
        \textbf{JiaT75} & tukaani-project/xz & \SI{10}{\months} & 24\\\hline
        Contributor C & linux-audit/audit-userspace & \SI{20}{\months} & 3\\\hline
		\bottomrule
	\end{tabular}
	\label{tab:resultsunreviewedselfmerged}
\end{table*}

The circumstances for each contributor in the results were investigated and are described briefly below:

\begin{itemize}
\item Contributor A had recently started actively contributing to the project and became the principal maintainer after an in-person meeting with the former maintainer. Contributor A has previously made commits with a corporate email address for a large American technology company and was verified as a current employee on the company’s website.
\item Contributor B was a long-term contributor to the project before its source code was hosted on GitHub. However, the email address on the Git identity used for their earlier contributions is not linked to their current GitHub user.
\item Contributor C uses a corporate email address for another sizable American technology company on their commits and has the same domain name as the project maintainer listed in the AUTHORS file.
\end{itemize}

Though none of the additional results were true positives, including the JiaT75 user and the overall limited number of matches proportional to the total number of users assessed (approximately 0.2\%), the query is suitable for detecting potential social engineering cases. Additionally, the discovery of earlier commits by the JiaT75 user suggests there is utility in using similar queries to investigate attacks that have already occurred.

\subsection{Querying for Significant Contribution Relative to Presence}

\subsubsection{Query Creation}

The second candidate criteria set to investigate earlier was a contributor making a significant share of all-time contributions relative to a contributor's presence in the repository history. 

As was done for the first candidate criteria, a query was prepared to collect the following values for each \textit{GithubUser} in each \textit{GithubRepository} where they had made at least one commit:

\begin{itemize}
\item The \textit{GitHubUser} username, creation timestamp, first authored commit timestamp, and all associated email addresses
\item The \textit{GithubRepository} name with its owner, creation timestamp, and first-authored commit timestamp
\item The difference between the \textit{GithubUser} and \textit{GithubRepository} first authored commit timestamps (i.e., the repository age at the time of the first contribution)
\item The difference between the current date and the \textit{GithubRepository} first authored commit timestamp (i.e., the repository age)
\item The percentage of commits authored by the user in the Git repository
\item The difference between the first and last user-authored commit timestamps, as a percentage of the difference of the first and last authored commit timestamps for the repository (i.e., what percentage of the repository history the contributor has been present)
\end{itemize}

This query returned 2023 unique combinations of users with contributions in repositories. The data for JiaT75 indicates they authored 17.4\% of commits to the repository, despite their contributions only taking place over 12.5\% of the repository's history. As the intent of the role is to catch potential social engineering efforts before they occur, looser thresholds were chosen, matching contributors who have authored more than 5\% of commits with less than 20\% presence in the repository. Rerunning the query with the threshold returned six unique contributors, and after applying the same exclusion for the first-authored over four years ago, was reduced to 4.

\subsubsection{Discussion of Results}

Table~\ref{tab:resultspresence} presents the relevant data from the results after executing the final query. The full query used is available in the supplementary material.

\begin{table*}
	\caption{Results of the Significant Contribution Relative to Presence Query}
	\centering
	\begin{tabular}{
        l
        l
        S[table-format=2.2]
        S[table-format=2.2]
    }
		\toprule
        Name & Repository & {\% of authored commits} & {\% of time presence in the commit history}
        \\\hline
        \midrule
        Contributor A & PCRE2Project/pcre2 & 6.91\% & 4.66\%\\\hline
        \textbf{JiaT75} & tukaani-project/xz & 17.38\% & 12.47\%\\\hline
        Contributor D & p11-glue/p11-kit & 7.18\% & 16.67\%\\\hline
        Contributor E & shadow-maint/shadow & 18.4\% & 18.28\%\\\hline
		\bottomrule
	\end{tabular}
	\label{tab:resultspresence}
\end{table*}

Again, the circumstances of each additional contributor were investigated, and are briefly described below:

\begin{itemize}
\item Contributor A also appeared in the results of the other query and was previously discussed.
\item Contributor D uses a corporate email address for the same domain name as Contributors B and C. They are also listed as a maintainer in the project README.md file.
\item Contributor E uses an email address with a domain associated with a sizable open-source software project and was verified as a long-term contributor through information published on the project's website. They are also listed as a project maintainer in the AUTHORS file.
\end{itemize}

Though once again the results did not contain any new true positives, the limited number of results, even a relatively relaxed threshold relative to what would have been observable for the JiaT75 persona, supports this also being a viable signal to investigate potential social engineering.

\section{Recommendations and Implications}

\subsection{Recommendations}

This research shows that social engineering attacks against open-source projects can potentially be detected by analysis of contributor metadata. Security practitioners aiming to monitor or investigate social engineering attacks against open-source projects should use automation where possible to collect and analyze contribution data from project contributors.

\subsection{Non-recommendations}

As was observed with the case of Contributor A, despite the aggregated data signaling suspicious or otherwise unusual behavior, upon investigation, it was discovered that they had been vetted by the former maintainer and recently taken over the responsibilities of the project.

Legitimate reasons exist for open-source contributors to use personas/aliases, or otherwise have a limited web presence outside of their open-source contributions. Security practitioners and project maintainers responding to signals raised by this kind of analysis should approach investigations without presumptions of malicious intent. While it is encouraged to be conscious and apply caution when granting privileges or transferring responsibilities to new parties, such caution should not dissuade or hinder legitimate contributors from participating. 

\subsection{Future Research}

This research validates the potential for investigation, but only on a smaller sample set of projects at a particular point in time. Potential future research could encompass one or more of the following elements:

\begin{itemize}
\item Continuous collection and ingestion of data.
\item Deeper traversal/further collection of data from the GitHub GraphQL API.
\item Additional data sources like the \textit{gharchive} project should be used to minimize the number of requests necessary for the GitHub GraphQL API.
\item Collection of data from other source code forges like GitLab.com or Gitea.com, as well as project-managed self-hosted instances.
\item Collection of data from mailing lists relevant to the project (for example, the project mailing list or Linux distribution mailing lists relevant to the project), and identification of similar usage of sock puppets to pressure project maintainers.
\item Collection of other web presence or social media activity for monitored contributing personas.
\item Use of LLM to identify major events like transferring maintainership in project issue trackers and mailing lists.
\item Use of sentiment analysis to identify unusual pressure applied towards maintainers in project issue trackers and mailing lists.
\end{itemize}

\section{Conclusion}

This research demonstrated the potential of using anomalous contribution patterns in open-source projects as a signal of social engineering attacks like the XZ Utils backdoor. Both sets of criteria tested against aggregate contributor data --- the timescale over which contributors gain privileged access to or become significant participants --- show viability as good signals for similar projects. Further work to continuously monitor projects and collect additional data sources is encouraged and could raise an early warning for a similar attack in the future.

\bibliographystyle{IEEEtran}
\bibliography{references}

\begin{IEEEbiography}[{\includegraphics[width=1in,height=1.25in,clip,keepaspectratio]{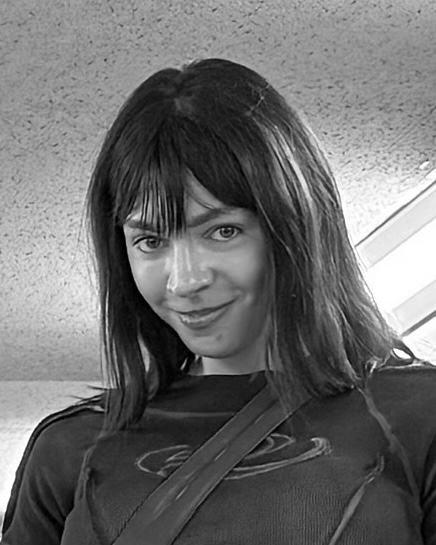}}]{Ruby Nealon} received an M.S. degree in information security engineering from SANS Technology Institute in 2025. She is currently a security engineer at GitLab, where she works on red team and platform safety operations.
\end{IEEEbiography}

\vfill

\end{document}